\documentstyle[12pt,fleqn]{article}
\begin{document}
\baselineskip=7.3mm

\begin{center}
{\large\bf Gamma-Ray Burst Afterglows: Effects of Radiative Corrections and
           Nonuniformity of the Surrounding Medium}

\vspace{5.0mm}
Z. G. Dai and T. Lu

{\em Department of Astronomy, Nanjing University, Nanjing 210093, China}
\end{center}

\vspace{10.0mm}

\begin{center}
ABSTRACT
\end{center}

The afterglow of a gamma-ray burst (GRB) is commonly thought to be due to
continuous deceleration of a relativistically expanding fireball in the
surrounding medium. Assuming that the expansion of the fireball is
adiabatic and that the density of the medium is a power-law function 
of shock radius, viz., $n_{\rm ext}\propto R^{-k}$,
we analytically study the effects of the first-order radiative
correction and the nonuniformity of the medium on a GRB afterglow.
We first derive a new relation among the observed time, the shock
radius and the fireball's Lorentz factor: $t_\oplus=R/4(4-k)\gamma^2c$, and
also derive a new relation among the comoving time, the shock radius and
the fireball's Lorentz factor: $t_{\rm co}=2R/(5-k)\gamma c$. 
We next study the evolution of the fireball by using the analytic solution of
Blandford and McKee (1976). The radiation losses may not significantly
influence this evolution. We further derive new scaling laws both
between the X-ray flux and observed time and between the optical 
flux and observed time. We use these scaling laws to 
discuss the afterglows of GRB 970228 and GRB 970616, and 
find that if the spectral index of the electron distribution is $p=2.5$,
implied from the spectra of GRBs, the X-ray afterglow of GRB970616 is well
fitted by assuming $k=2$.

\noindent
{\em Subject headings:} gamma-ray : bursts --- radiation
                        mechanisms : nonthermal

\newpage

\begin{center}
1. INTRODUCTION
\end{center}

The popular theoretical explanation for cosmological
gamma-ray bursts (GRBs) is based on the fireball model.
In this model, a GRB is thought to result from the dissipation of the
kinetic energy of a relativistically expanding fireball. This dissipation
can be either due to internal shocks formed during the collision between
the shells with different Lorentz factors in the fireball
(Rees \& M\'esz\'aros 1994; Paczy\'nski \& Xu 1994;
Sari \& Piran 1997), or due to external shocks formed by the fireball
colliding with the surrounding medium (Rees \& M\'esz\'aros 1992;
M\'esz\'aros \& Rees 1993; Katz 1994; Sari, Narayan, \& Piran 1996).
After the main GRB event, the expanding fireball is predicted to produce
delayed emission at longer wavelengths (Paczy\'nski \& Rhoads 1993;
Katz 1994; M\'esz\'aros \& Rees 1997; Wijers, Rees, \& M\'esz\'aros 1997;
Reichart 1997; Waxman 1997a,b; Vietri 1997a,b; Tavani 1997; Sari 1997).
It is easily understood that such an X, optical and/or radio afterglow
is in fact due to continuous deceleration of the expanding fireball.

Fortunately, the afterglows of GRBs have been detected recently in the
error boxes at the sites of seven GRBs, e.g. GRB970228, GRB970402,
GRB970508, GRB970616, GRB970815, GRB970828 and GRB971214.
The current fireball model to explain these afterglows has in fact been
divided into two sub-models. In the first sub-model, the fireball
expansion following a GRB is thought to be adiabatic (Wijers et al. 1997;
Reichart 1997; Waxman 1997a,b). This is a reasonable assumption
if the timescale for cooling of the accelerated electrons behind
the shocks rapidly becomes longer than the expansion timescale of
the fireball. The adiabatic expansion model has
given a scaling relation between the expansion Lorentz factor and
observed time: $\gamma\propto t_\oplus^{-3/8}$, provided that the
surrounding medium is uniform and the expansion is ultra-relativistic.
The studies of Wijers et al. (1997), Reichart (1997) and Waxman (1997a,b)
showed that the adiabatic expansion model may satisfactorily explain the
long-term behavior of the afterglows of GRB970228 and GRB970508. The more
detailed numerical calculations by Huang et al. (1997) are consistent with
these studies. An alternative sub-model accounted possibly for several
properties of the afterglows of these two bursts has been presented by
Vietri (1997b), who postulated that the expansion is highly radiative. This
requires that the accelerated electrons always cool more rapidly than
the fireball expands.

The purpose of the present work is to study analytically the effects of
radiative corrections and nonuniformity of the surrounding matter on a GRB
afterglow based on the adiabatic expansion model. Our study is stimulated
by two motivations. First, Sari (1997) recently found that radiation losses
may significantly influence the hydrodynamical evolution of the fireball
in the uniform medium. Second, it is possible that the sources of GRBs are
merging of neutron star binaries (Narayan et al. 1992; Vietri 1996), failed
supernovae (Woosley 1993), accretion-induced phase transitions of neutron
stars (Cheng \& Dai 1996), or hypernovae (Paczy\'nski 1998). If so, one may
not think that the surrounding matter is uniform. A massive star as
a progenitor of a GRB has possibly produced a stellar wind or a supernova
remnant exists around a neutron star which is a source of the GRB. 
The stellar wind and/or the supernova remnant constitute the nonuniform
surrounding medium of the GRB. It can be expected that the nonuniformity
of the medium can shorten or prolong the relativistic expansion of the
fireball. Such a behaviour should be compared with observations
of afterglows of GRBs.

In order to reach the purpose of the present work, in the next section
we first derive a new relation among the observed time, the shock
radius and the fireball's Lorentz factor, assuming that the density of
the surrounding medium is a power-law function of shock radius.
We also obtain a new relation among the comoving time, the shock
radius and the fireball's Lorentz factor. We show that, in the case of
uniform medium, the former relation turns out ot be consistent with the
expression derived by Sari (1997), but the latter relation is different
from the usual expression by a factor of 5/2. We next study the evolution
of the relativistic fireball in the nonuniform medium by using the analytic
solution found by Blandford \& McKee (1976), and investigate the
effect of the first-order radiative correction on the evolution. We
find this effect may be insignificant. In section 3, we derive the
X-ray flux from the fireball as a function of observed time,
and give a scaling relation between the optical flux and observed time.
In section 4, we use our model to discuss the afterglows of GRB970228 and
GRB970616. In the final section, we make a summary for our work.

\begin{center}
2. THREE MEASURES OF TIME AND FIRST-ORDER RADIATIVE CORRECTION
\end{center}

Even though the source of GRBs is unknown, one believes the following
scenario for GRBs: A compact source ($\sim 10^7\,$cm) releases an energy
$E$ comparable to that observed in gamma rays, $E\sim 10^{51}\,$erg, over
a time less than 100s. The huge optical depth in the source results in
an initial fireball that expands and accelerates to relativistic velocity.
During the acceleration, the fireball energy is converted to bulk kinetic
energy with a Lorentz factor $\eta\sim 300$. Of course, internal shocks
may be formed in this period. Subsequently, the expansion of the fireball
starts to be significantly influenced by the swept-up medium and
two external shocks are formed: a forward blast wave and a reverse shock
(Rees \& M\'esz\'aros 1992). A GRB may be produced by nonthermal processes
in shocks. After the GRB, the relativistic blast wave continues to
sweep up the medium. The fireball may go into an adiabatic expansion phase.
For simplicity, we assume that the medium density varies with shock radius
based on the following expression:
\begin{equation}
n_{\rm ext}=n_0\left(\frac{R}{R_0}\right)^{-k}
\,\,\,\, {\rm for} \,\, R\ge R_0\,,
\end{equation}
where $n_0$ is the medium density at the radius $R_0$ at which the fireball
starts to be significantly influenced by the medium. As usual, $R_0$
is defined as
\begin{equation}
R_0=\left(\frac{3E}{4\pi n_0m_pc^2\eta^2}\right)^{1/3}
=10^{16}E_{51}^{1/3}n_0^{-1/3}\eta_{300}^{-2/3}\,\,\,{\rm cm}\,,
\end{equation}
where $E_{51}=E/10^{51}\,{\rm erg}$, $\eta_{300}=\eta/300$,
and $n_0$ is in units of 1\,cm$^{-3}$.

The use of a power-law dependence of the density surrounding a GRB source
is motivated by the following points. First, some theoretical studies
of supervova remnants (for a review see Ostriker \& McKee 1988)
suggest a power-law dependence of the remnant density
at radius larger than some value. Second, the actual dependence might be more
complicated than a power-law form. However, only in the case with a
power-law form can we analytically study the evolution of a relativistic
fireball. A study of the relatvistic-fireball expansion in the nonuniform
medium without a power-law dependence of the matter density on radius
needs numerical simulations and thus is beyond the scope of this paper.

\begin{center}
{\em 2.1. Three Measures of Time}
\end{center}

We assume that the expansion is ultra-relativistic, and radiation
losses are small. This implies that when most of the energy has been given
to the medium, the energy in the shocked medium is constant and
approximately equal to $E$. The rest mass of the shocked medium
$M\propto R^{3-k}$. Since the medium was thermalized by the relativistic
blast wave, its energy in the observer's frame is $\sim M\gamma^2\sim E$,
where $\gamma$ is the Lorentz factor of the shocked medium just behind
the shock. Thus, we have the following scaling law
\begin{equation}
\gamma\propto R^{-(3-k)/2}\,.
\end{equation}
It should be emphasized that for a relativistic strong blast wave the
shock's Lorentz factor $\Gamma=\sqrt{2}\gamma$ (Blandford \& McKee 1976).

According to the scaling law (3), we next investigate relations among
three different measures of time denoted as $t$, $t_{\rm co}$ and $t_\oplus$
which are measured in the rest frame of the burster, in the frame comoving
with the fireball, and in the observer's frame respectively. When the shock
propagates a small distance $\delta R\approx c\delta t$, photons that
are emitted from the shock will be observed on the observed timescale
of $\delta t_\oplus=\delta R/(2\Gamma^2c)=\delta R/(4\gamma^2c)$ while
the change in time in the frame comoving with the fireball is $\delta
t_{\rm co}=\delta R/(\gamma c)$. Integrating these equations over time and
using the scaling law (3), we obtain
\begin{equation}
t_\oplus=\frac{R}{4(4-k)\gamma^2c}=\frac{t}{4(4-k)\gamma^2}\,,
\end{equation}
and
\begin{equation}
t_{\rm co}=\frac{2R}{(5-k)\gamma c}=\frac{2t}{(5-k)\gamma}\,.
\end{equation}
In the case of uniform medium, viz., $k=0$, eq. (4) turns out to be
consistent with the expression derived by Sari (1997). In this case,
comparing eq. (5) with the commonly used expression $R/(\gamma c)$
(Waxman 1997a,b; Vietri 1997b), we see the present expression is
a factor of 5/2 smaller.

\begin{center}
{\em 2.2. First-Order Radiative Correction}
\end{center}

After having eqs. (4) and (5), we can study the adiabatic evolution
of the fireball in the nonuniform medium, and investigate
the effect of the first-order radiative correction on the evolution.

The starting point of our study is eq. (69) of Blandford \& McKee (1976)
about the total energy of the fireball:
\begin{equation}
E=\frac{16\pi n_{\rm ext}m_pc^2\gamma^2R^3}{17-4k}\,,
\end{equation}
which is constant (to lowest order). Here $n_{\rm ext}$ is to be interpreted
as the external density at the position of the shock. Combining eq. (6) with
eqs. (1) and (4), we get
\begin{equation}
\gamma=\gamma_0\left(\frac{t_\oplus}{t_0}\right)^{-(3-k)/(8-2k)}\,,
\end{equation}
where $\gamma_0$ and $t_0$ are defined as
\begin{equation}
\gamma_0=\left[\frac{(17-4k)E}{16\pi n_0m_pc^2R_0^3}\right]^{1/(8-2k)}\,,
\end{equation}
and
\begin{equation}
t_0=\frac{R_0}{4(4-k)c}\,.
\end{equation}
Accordingly, the shock radius can be written as
\begin{equation}
R=R_0\gamma_0^2\left(\frac{t_\oplus}{t_0}\right)^{1/(4-k)}\,.
\end{equation}
It is easily seen that in the case of $k=0$ eqs. (7) and (10) are in
agreement with those derived by Sari (1997).

In the following we consider only synchrotron emission from the accelerated
electrons. We neglect the contribution of inverse-Compton (IC) emission from
these electrons. This is becasue IC emission is not of importance
particularly at late times of the fireball expansion (for a discussion
to be seen in the final paragraph of this subsection).
In order to calculate the effect of synchrotron emission, we need to
determine the magnetic field strength and electron energy. As usual, we
assume that the magnetic energy density in the comoving frame is a fraction
$\xi_B$ of the total thermal energy density $e^\prime=4\gamma^2n_{\rm ext}
m_pc^2$, viz., $B^\prime=(8\pi \xi_Be^\prime)^{1/2}$, and that
the electrons carry a fraction $\xi_e$ of the energy. This implies
that the Lorentz factor of the random motion of a typical electron
in the comoving frame is $\gamma_{\rm em}=\xi_e\gamma m_p/m_e$. The
ratio of the comoving-frame expansion time, $t_{\rm co}=2R/(5-k)\gamma c$,
to the synchrotron cooling time, $t_{\rm syn}=6\pi m_ec/\sigma_T\gamma_
{\rm em}{B^\prime}^2$, is
\begin{equation}
\frac{t_{\rm co}}{t_{\rm syn}}=\left(\frac{t_1}{t_\oplus}\right)^{2/(4-k)}\,,
\end{equation}
where $t_1$ has been defined based on
\begin{equation}
t_1^{2/(4-k)}=\frac{128(4-k)}{3(5-k)}\left(\frac{m_p}{m_e}\right)^2
\xi_e\xi_B(\sigma_Tn_0ct_0)\gamma_0^{3-2k}t_0^{2/(4-k)}\,.
\end{equation}
Thus, we can define the radiative efficiency as
\begin{equation}
\xi=\frac{t_{\rm syn}^{-1}}{t_{\rm co}^{-1}+t_{\rm syn}^{-1}}
=\frac{(t_1/t_\oplus)^{2/(4-k)}}{(t_1/t_\oplus)^{2/(4-k)}+1}\,.
\end{equation}
This parameter accounts for a fraction of the energy of the accelerated
electrons which is radiated away by the synchrotron emission. For
$t_\oplus\ll t_1$, $\xi\approx 1$; but for $t_\oplus\gg t_1$, $\xi\approx
(t_1/t_\oplus)^{2/(4-k)}$. Following eq. (84) of Blandford \& McKee (1976),
we derive the energy loss rate during the deceleration which is given by
$4^4(4-k)^2\pi \gamma^8c^3t_\oplus^2n_{\rm ext}m_pc^2$ multiplied by
$\xi_e\xi$, where eq. (4) has been used. Using eqs. (7)-(9),
we get the total power radiated per unit time:
\begin{equation}
\frac{dE}{dt_\oplus}=-\xi_e\xi\frac{17-4k}{4(4-k)}\frac{E}{t_\oplus}\,.
\end{equation}
Integrating this equation over time, we further obtain
\begin{equation}
E(t_\oplus)=E_0\left[\frac{(t_1/t_\oplus)^{2/(4-k)}+1}
{(t_1/t_{\rm in})^{2/(4-k)}+1}\right]^{\frac{17-4k}{8}\xi_e}\,,
\end{equation}
where $t_{\rm in}$ is the observed initial time of the afterglow
which is of the order of magnitude $\sim 10\,$s.
We now define the parameter $f$ as
\begin{equation}
f=\left[\left(\frac{t_1}{t_{\rm in}}\right)^{2/(4-k)}+1\right]
^{-\frac{17-4k}{8}\xi_e}\,.
\end{equation}
This parameter is in fact the ratio of the fireball energy at
$t_\oplus\gg t_1$ to the initial total energy, and thus accounts for
the effect of the first-order radiative correction. To estimate
$f$, we adopt the following values: $E_{51}=4$, $\eta_{300}=1$,
$n_0=1\,{\rm cm}^{-3}$, $\xi_B=0.1$, and $\xi_e\sim 0.1$--$0.3$. Table 1
gives the values of $\gamma_0$, $t_0$, $t_1$ and $f$ for different $k$.
It can be seen from this table that for $\xi_e\sim 0.1$ the effect
of the radiative corrections may be insignificant, and for $\xi_e\sim 0.3$
this effect may not be of importance in the case of $k>0$. This
conclusion disagrees with that of Sari (1997), who didn't consider
the factor $\xi$ in eq. (14). For this reason, we will not take into
account this effect in the next text.

In the remainder of this subsection, we want to discuss the validity
of two of our assumptions. First, the fireball expansion has been assumed
to be adiabatic. As described in the introduction, this assumption
in fact requires that the timescale for cooling of the accelerated electrons
behind the shocks rapidly becomes longer than the fireball expansion
timescale. Now we assume that the initial expansion of the fireball
is radiative. In this case (e.g., see Vietri 1997b), the relations among
the shock radius, comoving-frame time and observed time are
$R=4(7-2k)\gamma^2ct_\oplus$ and $R=(4-k)\gamma ct_{\rm co}$, and the
Lorentz factor of the fireball decreases as
\begin{equation}
\gamma=\eta\left(\frac{t_\oplus}{t_{\rm 0r}}\right)^{-\frac{3-k}{7-2k}},
\end{equation}
where
\begin{equation}
t_{\rm 0r}=\frac{R_0}{(7-2k)4\eta^2c}=\frac{1}{7-2k}E_{51}^{1/3}n_0^{-1/3}
\eta_{300}^{-8/3}\,\,\,\,{\rm s}.
\end{equation}
Thus, the ratio of the comoving-frame expansion timescale to the synchrotron
cooling timescale is given by
\begin{equation}
\frac{t_{\rm co}}{t_{\rm syn}}=\left(\frac{t_{\rm 1r}}{t_{\oplus}}\right)^
{\frac{5-k}{7-2k}},
\end{equation}
where $t_{\rm 1r}$ is defined through the following expression:
\begin{equation}
t_{\rm 1r}^{(5-k)/(7-2k)}=\frac{64(7-2k)}{3(4-k)}\left(\frac{m_p}{m_e}
\right)^2\xi_e\xi_B(\sigma_Tn_0ct_{\rm 0r})\eta^4t_{\rm 0r}^{(5-k)/(7-2k)}\,.
\end{equation}
The values of $t_{\rm 1r}$ for different $k$ are calculated in Table 2.
According to this table and eq. (19), we see that
$t_{\rm syn}>t_{\rm co}$ for $t_\oplus>t_{\rm 1r}$, and therefore conclude
that even if the fireball starts off with radiative dynamics, the transition
to adiabatic evolution comes soon after the GRB.

Second, we have not considered the contribution of IC emission to cooling
of the accelerated electrons. It is well known that whether IC emission
is important depends on the ratio of the IC power to synchrotron power:
\begin{equation}
y=\frac{e'_S}{e'_B}\,,
\end{equation}
where $e'_S$ and $e'_B$ are the synchrotron-photon and magnetic-field energy
densities. For $t_{\rm co}<t_{\rm syn}$ and when emission is dominated
by the synchrotron process, as argued by Waxman (1997a), the energy density
$e'_S$ is a fraction $t_{\rm co}/4t_{\rm syn}$ of the electron energy density,
and thus the ratio $y$ is given by
\begin{equation}
y=\frac{\xi_e}{\xi_B}\frac{t_{\rm co}}{4t_{\rm syn}}
=\left(\frac{t_2}{t_\oplus}\right)^{2/(4-k)}\,,
\end{equation}
where $t_2$ has been defined by
\begin{equation}
t_2=\left(\frac{\xi_e}{4\xi_B}\right)^{(4-k)/2}t_1\,.
\end{equation}
Therefore, we can conclude from Table 1 that IC emission is not an important
process for cooling of the accelerated electrons behind the shocks
for $t_\oplus>t_2$.

\begin{center}
3. THE X-RAY AND OPTICAL RADIATION
\end{center}

We first study the X-ray flux from a relativistic fireball expanding in the
nonuniform medium. The total power radiated per unit time has been
given by eq. (14). In order to calculate the X-ray flux,
we need to consider radiation mechanisms of the accelerated
electrons behind the shock. As shown in the last section, the synchrotron
emission is the main mechanism for cooling of these
electrons. We assume that the electron distribution behind the shock
is power-law:
\begin{equation}
\frac{dN_e}{d\gamma_e}\propto \gamma_e^{-p}\,,\,\,\,\,\,\,{\rm for}
\,\,\,\,\,\,\gamma_{\rm em}\le\gamma_e\le\gamma_{\rm e,max}\,.
\end{equation}
The estimate of $\gamma_{\rm e,max}$ can be obtained by equating
the electron acceleration timescale, $t_a=\gamma_em_ec/eB^\prime$,
to the synchrotron cooling timescale so that
\begin{equation}
\gamma_{\rm e,max}=\left(\frac{6\pi e}{\sigma_T B^\prime}\right)^{1/2}\,.
\end{equation}
It should be pointed out that
even though the maximum Lorentz factor estimated by this equation is
rather high, the effect of IC emission on $\gamma_{\rm e,max}$ can be
neglected, as seen in eq. (22).
For the electron distribution with eq. (24), the fraction of synchrotron
power radiated in the X-ray region is (Vietri 1997a,b)
\begin{equation}
f_x=\frac{\epsilon_u^{(3-p)/2}-\epsilon_l^{(3-p)/2}}{\epsilon_{\rm max}
^{(3-p)/2}}\,,
\end{equation}
where $\epsilon_u$ and $\epsilon_l$ are the upper and lower limits
of the BeppoSAX instruments, 2 and 10keV respectively, and
$\epsilon_{\rm max}$ is given by
\begin{equation}
\epsilon_{\rm max}=\frac{\hbar eB^\prime}{m_ec}\gamma_{\rm e,max}^2\gamma
=160\gamma\,\,\,{\rm MeV}\,.
\end{equation}
Inserting eq. (27) into eq. (26), we find
\begin{equation}
f_x\approx [6\times 10^{-5}/\gamma]^{(3-p)/2}\,,
\end{equation}
where the number in the brackets is two orders smaller than that of
Vietri (1997b). Therefore, the expected X-ray flux is
\begin{equation}
F_x=\frac{dE}{dt_\oplus}\frac{f_x}{4\pi d^2}
=1.0\times 10^{-6}\,{\rm erg}\,{\rm s}^{-1}\,{\rm cm}^{-2}\,
\left(\frac{d}{1{\rm Gpc}}\right)^{-2}\left(\frac{t_\oplus}{1\,{\rm s}}
\right)^{-a}\,,
\end{equation}
where $d$ is the source distance, the constant has been computed for $k=0$,
$p=2.5$, $\xi_e=0.1$, and the values of the other
parameters used in Table 1. For $t_\oplus\ll t_1$ in the above equation,
\begin{equation}
a=1-\frac{3-k}{8-2k}\frac{3-p}{2}\,;
\end{equation}
but for $t_\oplus \gg t_1$,
\begin{equation}
a=\frac{6-k}{4-k}-\frac{3-k}{8-2k}\frac{3-p}{2}\,.
\end{equation}

We next discuss the optical flux from the accelerated electrons
behind the shock. For the electron distribution with eq. (24),
the observed frequency of synchrotron emission at peak flux
is
\begin{eqnarray}
\nu_m & = & \frac{3}{4\pi}\gamma\gamma_{\rm em}^2\frac{eB^\prime}{m_ec}
=\frac{6}{\sqrt{2\pi}}\xi_e^2\xi_B^{1/2}\gamma_0^{4-k}
\left(\frac{m_p}{m_e}\right)^2\frac{e(n_0m_pc^2)^{1/2}}{m_ec}
\left(\frac{t_\oplus}{t_0}\right)^{-3/2}  \nonumber \\
& = & 8.0\times 10^{21}\xi_e^2\xi_B^{1/2}E_{51}^{1/2}\left(\frac{17-4k}{17}
\right)^{1/2}\left(\frac{4}{4-k}\right)^{3/2}
\left(\frac{t_\oplus}{1\,{\rm s}}\right)^{-3/2}\,{\rm Hz}\,,
\end{eqnarray}
where eqs. (8) and (9) have been used. Eq. (32) shows that $\nu_m$
is weakly dependent of $k$. The typical spectrum of the synchrotron
emission has the following form: $F_\nu\propto \nu^\alpha$, where
$\alpha=1/3$ for $\nu<\nu_m$, and $\alpha=-(p-1)/2$ for $\nu>\nu_m$.
Since the comoving electron density $n_e^\prime\propto \gamma n_{\rm ext}
\propto t_\oplus^{-(3+k)/(8-2k)}$, and the comoving width of the fireball
$\Delta R'\propto R/\gamma\propto t_\oplus^{(5-k)/(8-2k)}$, then
the comoving intensity $I_\nu^\prime\propto n_e^\prime B^\prime \Delta R'
\propto t_\oplus^{-(1+2k)/(8-2k)}$ (M\'esz\'aros \& Rees 1997;
Wijers et al. 1997). Thus, the observed peak flux as a function of time is
$F_{\nu_m}\propto t_\oplus^2\gamma^5I_{\nu_m}^\prime
\propto t_\oplus^{-k/(8-2k)}$. Once $\nu_m$ has entered the optical
region, the observed flux at any frequency must vary according to
\begin{equation}
F_\nu=F_{\nu_m}(\nu/\nu_m)^\alpha\propto t_\oplus^{-b}\,,
\end{equation}
where for $\nu<\nu_m$,
\begin{equation}
b=-\frac{2-k}{4-k}\,;
\end{equation}
for $\nu>\nu_m$,
\begin{equation}
b=\frac{k}{8-2k}+\frac{3(p-1)}{4}\,.
\end{equation}
It can be seen from eqs. (33)-(35) that only for $k<2$ the observed optical
flux first increases and then decreases, but for $k=2$ the flux is first
kept to be constant and subsequently declines.

\begin{center}
4. DISCUSSION
\end{center}

We first use our model to discuss the X-ray abd optical afterglow of
GRB970228. According to the observational results summarized by
Wijers et al. (1997), we find $p\approx 2.4$ and $k\approx 0$ by solving
eqs. (31) and (35), which is well consistent with that of Waxman (1997a).
The result of $p\approx 2.4$ is consistent with the mean
spectral index ($p=2.5$) of GRBs measured in Band et al. (1993).
This shows that the spectral index of the electron distribution due to shock
acceleration is likely to be similar for GRBs and their afterglows.

Our model can also be applied to discussing the X-ray afterglow of GRB970616.
This burst was detected by BATSE on June 16.757UT. About 20min after
the initial trigger, a transient X-ray XTE source was found in the
error box of this burst (Connaughton et al. 1997), and 4 hours following
the burst, scanning observations with the Proportional Counter Array on the
{\em Rossi X-ray Timing Explorer} (RXTE) revealed an X-ray afterglow
in the band 2-10keV with a flux $\sim 1.1\times 10^{-11}\,{\rm erg}\,{\rm
s}^{-1}\,{\rm cm}^{-2}$ (Marshall et al. 1997). On June 20.35UT, {\em ASCA}
detected an X-ray flux from the XTE/IPN error box of GRB970616 with
$\sim 3.7\times 10^{-14}\,{\rm erg}\,{\rm s}^{-1}\,{\rm cm}^{-2}$ in
the band 0.7-7keV (Murakami et al. 1997). Using these values of the X-ray
flux in eq. (29), we get $a\approx 1.86$. After knowing the value of
$a$ and assuming that $p$ is equal to the mean spectral index of GRBs
measured in Band et al. (1993), that is, $p\approx 2.5$, we solve
eq. (35) and find $k=2$. This result implies nonuniformity
of the surrounding medium.

We have found that the afterglows of GRB970616 and GRB970228 are
well explained by assuming two cases with $k=2$ and 0 respectively.
We easily understand these two cases. In the first case,
a neutron star as the GRB source has lain in a supernova remnant and/or
a stellar wind due to the star's low velocity, so the postburst fireball
has expanded in such a nonuniform medium; but in the second case,
we conjecture that since the velocity of a neutron star as the progenitor of
the GRB was very high, the GRB source has left a supernova remnant and/or
a stellar wind, and the fireball has met the uniform interstellar medium.

\begin{center}
5. SUMMARY
\end{center}

A GRB has been commonly believed to result from the dissipation of the
kinetic energy of a relativistically expanding fireball, and its X, optical 
and/or radio afterglow is due to continuous deceleration 
of the fireball. In this paper, we have assumed that the 
expansion of the fireball is adiabatic and ultra-relativistic. If 
compact objects (neutron stars or black holes) are an origin
of the GRB, the surrounding medium of the fireball may be nonuniform 
due to the existence of a stellar wind and/or a supernova remnant.
For simplicity, we have assumed that the density of the medium 
is a power-law function of shock radius, viz., 
$n_{\rm ext}\propto R^{-k}$. In addition, radiation losses
may significantly influence the hydrodynamical evolution of the 
fireball (Sari 1997). In view of these two important arguments, we have
analytically studied the effects of the first-order radiative correction and
the nonuniformity of the medium on the GRB afterglow in this paper.
The results of our study are summarized as follows:

First, we have derived a new relation among the observed time, the shock
radius and the fireball's Lorentz factor. We have also obtained 
a new relation among the comoving time, the shock radius and the 
fireball's Lorentz factor. We have shown that, in the case of
uniform medium, the former relation turns out ot be consistent with the
expression derived by Sari (1997), but the latter relation is smaller than
the usually used expression by a factor of 5/2. 

Second, we have used the analytic solution of Blandford \& McKee (1976)
to derive the fireball's Lorentz factor and the shock
radius as functions of observed time, which show that the nonuniformity
of the medium must shorten ($k<0$) or prolong ($k>0$)
the relativistic expansion of the fireball. Using these functions, we have
further derived the radiation energy loss rate, and found the first-order
radiative correction may be insignificant. This conclusion disagrees
with that of Sari (1997) who neglected the radiative efficiency 
defined in eq. (13). 

Third, we have derived new scaling laws both between the X-ray flux
and observed time and between the optical flux and observed 
time. We have found that only for $k<2$ the observed optical
flux first increases and then decreases, but for $k=2$ the flux is first
kept to be constant and subsequently declines.

Finally, We have used our model to 
discuss the afterglows of GRB 970616 and GRB 970228.
We have seen that the afterglow of GRB970616 is well fitted by assuming 
$k=2$. This value implies of the nonuniformity of 
the medium.

\vspace{5mm}
We would like to thank the referee, Dr. Ralph Wijers, for very valuable 
suggestions, and Drs. K. S. Cheng and D. M. Wei for many helpful
discussions. This work was supported by the National Natural Science
Foundation of China.

\newpage
\baselineskip=4mm

\begin{center}
REFERENCES
\end{center}

\begin{description}
\item Band, D. et al. 1993, ApJ, 413, 281
\item Blandford, R. D., \& McKee, C. F. 1976, Phys. Fluids, 19, 1130
\item Cheng, K. S., \& Dai, Z. G. 1996, Phys. Rev. Lett., 77, 1210
\item Connaughton, V. et al. 1997, IAU Circular No. 6683
\item Huang, Y. F., Dai, Z. G., Wei, D. M., \& Lu, T. 1997, MNRAS, in press
\item Katz, J. 1994, ApJ, 422, 248
\item Marshall, F. E. et al. 1997, IAU Circular No. 6683
\item M\'esz\'aros, P., \& Rees, M. J. 1993, ApJ, 405, 278
\item ---------------. 1997, ApJ, 476, 232
\item Murakami, T. et al. 1997, IAU Circular No. 6687
\item Narayan, R., Paczy\'nski, B., \& Piran, T. 1992, ApJ, 395, L83
\item Ostriker, J. P., \& McKee, C. F. 1988, Rev. Mod. Phys., 60, 1
\item Paczy\'nski, B. 1998, ApJ, 494, L45
\item Paczy\'nski, B., \& Rhoads, J. 1993, ApJ, 418, L5
\item Paczy\'nski, B., \& Xu, G. 1994, ApJ, 427, 708
\item Rees, M. J., \& M\'esz\'aros, P. 1992, MNRAS, 258, 41p
\item ---------------. 1994, ApJ, 430, L93
\item Reichart, D. E. 1997, ApJ, 485, L57  
\item Sari, R. 1997, ApJ, 489, L37
\item Sari, R., Narayan, R., \& Piran, T. 1996, ApJ, 473, 204
\item Sari, R., \& Piran, T. 1997, ApJ, 495, 270
\item Tavani, M. 1997, ApJ, 483, L87
\item Vietri, M. 1996, ApJ, 471, L91
\item ---------------. 1997a, ApJ, 478, L9
\item ---------------. 1997b, ApJ, 488, L105
\item Waxman, E. 1997a, ApJ, 485, L5
\item ---------------. 1997b, ApJ, 489, L33
\item Wijers, R., Rees, M. J., \& M\'esz\'aros, P. 1997, MNRAS, 288, L51
\item Woosley, S. 1993, ApJ, 405, 273
\end{description}

\newpage
\baselineskip=6.5mm

\begin{table}{Table 1. The typical values of some parameters for 
                       different $k$ \\ in the case of adiabatic evolution.
                       \\ \\}
\begin{tabular}{ccccccccc}
\hline\hline
~ & $\gamma_0$ & $t_0({\rm s})$ & $t_{1,{\rm a}}({\rm s})$ & $t_{1,{\rm b}}
({\rm s})$ & $f_{\rm a}$ & $f_{\rm b}$ & $t_{2,{\rm a}}$(s)
& $t_{2,{\rm b}}(s)$ \\ \hline
$k=0$ ........ & 4.67 & $3.31\times 10^4$ & 200 & 1800 & 0.71 & 0.20 & 12.5 &
               1013 \\
$k=1$ ........ & 7.46 & $4.41\times 10^4$ & 30 & 140 & 0.81 & 0.42 & 3.75 &
               90.9  \\
$k=2$ ........ & 18.56 & $6.61\times 10^4$ & 5.0 & 15.0 & 0.94 & 0.73 & 1.25
               & 11.3  \\
\hline\hline
\end{tabular}
\end{table}
\noindent
Notes --- $\gamma_0$, $t_0$, $t_1$, $f$ and $t_2$ are defined by eqs.
       (8), (9), (12), (16) and (23) \\ in the text, respectively. 
       The values are computed for $E_{51}=4$,  $\eta_{300}=1$, \\
       $n_0=1\,{\rm cm}^{-3}$ and $\xi_B=0.1$. The subscripts
       "a" and "b" represent \\ $\xi_e=0.1$ and 0.3 respectively.

\newpage
\begin{table}{Table 2. The typical values of some parameters for
                       different $k$ \\ in the case of radiative evolution.
                       \\ \\}
\begin{tabular}{cccc}
\hline\hline
~ & $t_{\rm 0r}{\rm (s)}$ & $t_{\rm 1r,a}{\rm (s)}$ & $t_{\rm 1r,b}{\rm (s)}$
 \\ \hline
$k=0$ ............... & 0.227 & 47.3 & 220 \\
$k=1$ ............... & 0.317 & 53.2 & 210 \\
$k=2$ ............... & 0.529 & 47.9 & 144 \\
\hline\hline
\end{tabular}
\end{table}
\noindent
Notes --- $t_{\rm 0r}$ and $t_{\rm 1r}$ are defined by eqs.
       (18) and (20) in the text, \\ respectively. The values
       of the parameters of the fireball and medium \\ are taken as 
       in Table 1. The subscripts "a" and "b" represent 
       \\ $\xi_e=0.1$ and 0.3 respectively.

\end{document}